\documentstyle[twocolumn,prl,aps,epsf,graphicx]{revtex}
\begin{document}
\draft
\twocolumn[
\hsize\textwidth\columnwidth\hsize\csname @twocolumnfalse\endcsname

\draft
%\title{Off-Equilibrium Dynamics in the Energy Landscape of\
% a Simple Model Glass}
%\title{Role of Saddles and Minima in the Off-Equilibrium Dynamics\ 
%of Simple Model Glass}
%\title{Role of Saddles and Minima in the Dynamics of an Aging Model Glass}
\title{Topological Description of the Aging Dynamics in Simple Glasses}
\author{ 
        L.~Angelani$^{1}$,
        R.~Di Leonardo$^{2}$,
        G.~Parisi$^{3}$, and
        G.~Ruocco$^{3}$
        }
\address{ 
         $^1$
         Universit\'a di Trento and INFM, I-38050, Povo, Trento, Italy.\\
         $^2$
         Universit\'a di L'Aquila
         and INFM,
         I-67100, L'Aquila, Italy. \\
         $^3$
         Universit\'a di Roma {\rm La Sapienza} and INFM, I-00185,
          Roma, Italy.
        }
\date{\today}
\maketitle
\begin{abstract}
We numerically investigate the aging dynamics of a monatomic 
Lennard-Jones glass, focusing on the topology of the potential 
energy landscape which, to this aim, has been partitioned in basins 
of attraction of stationary  points (saddles and minima). 
The analysis of the stationary points
visited during the aging dynamics shows the existence of two distinct regimes: 
{\it i}) at short times, $t<t_c$, the system visits basins of saddles
whose energies and orders decrease with $t$; {\it ii}) at long times, $t>t_c$,
the system  mainly lies in basins pertaining to minima of slowly 
decreasing energy.
%with $t$. 
The dynamics for $t>t_c$ can be represented by a 
simple random walk on a network of minima with a jump 
probability proportional to the inverse of the waiting time. 
%time elapsed  since the system has been driven out of equilibrium. 

\end{abstract}
]

%%%%%%%%%%%%%%%  TEXT  %%%%%%%%%%%%%%%%
The very slow approach to equilibrium of glassy systems 
is a very interesting and intriguing physical process 
that, in the last few years, has received considerable attention
from analytical, numerical and experimental point of view \cite{revaging}.
%When a system crosses a phase boundary towards a glassy state 
When a glassy system is suddenly brought in the glassy state from 
an high temperature (or low density) equilibrium state,
it starts to explore the phase space with an off-equilibrium
dynamics which slows down as the system ``ages''.
During this aging regime one time physical quantities, 
%such as energy, 
that are time independent on average at equilibrium, 
slowly approach their equilibrium values, 
while two time quantities,
like the correlation functions,
depend on both times due to the lack
of time translation invariance.
The study of the off-equilibrium dynamics of glassy systems, 
and in particular of spin-glasses, has allowed
to obtain important informations on the phase-space properties 
of the systems itself (behavior of the order parameter in the low 
temperature equilibrium phase \cite{cugli,framez,framepa,mari})
and to formulate off-equilibrium relations
generalizing the usual equilibrium ones (generalized fluctuation-dissipation
relation \cite{revaging,cugli}) with the introduction of an 
effective temperature \cite{kurpel}.
The investigation of the off-equilibrium properties seems 
to indicate the correctness of the conjecture stating 
the similarity of structural glasses with some spin glass model 
(spin-glasses with one step replica symmetry breaking) 
\cite{equiv1,equiv2,paout,baout,dileo}.
The way a glassy system goes towards equilibrium is then not only an 
interesting problem by itself but it is also of great relevance
in view of a deeper understanding of the nature of glass
transition. 
In this respect, many efforts have been devoted to the understanding of the role
played by the potential energy landscape in the equilibrium dynamics
of supercooled liquids \cite{pela1}.
The trajectory of the representative point in the $3N$ configuration 
space can be mapped into a sequence of locally stable points 
(the so called {\it Inherent Structures}, IS \cite{stilli}), that are the local minima of the 
total potential energy $V$: to each instantaneous configuration during the 
dynamical evolution of the system one can associate an IS by a steepest 
descent path in the $V$ surface. The properties of the IS has been found 
to be very useful to clarify many features in the dynamics and the thermodynamics
of supercooled liquids in \cite{sast,buec,scior,scho} and off \cite{sciortplot,sciortar} equilibrium.
An extension of such an analysis including all stationary points of $V$
(minima and saddles) \cite{cava0,ange,cava}
has been recently proposed to form a better representation
of the system during the equilibrium dynamics.
This approach gives a novel point of view on the equilibrium
dynamics of supercooled liquids based on a detailed  topological analysis 
of the potential energy landscape.
In particular it emerges \cite{ange} 
that the order of the sampled saddles during the equilibrium dynamics is a well defined
increasing function of temperature and it appears to vanish at the so-called 
mode-coupling \cite{mct} transition temperature $T_{MCT}$, which then marks 
the crossover from 
a dynamics between basins of saddles ($T > T_{MCT}$)
to a dynamics between basins of minima ($T < T_{MCT}$).
Similar results have been found \cite{donati} in the framework
of Instantaneous Normal Modes analysis \cite{keyes}.

In this Letter we numerically investigate the off-equilibrium
dynamics of a simple model glass, by mapping the true dynamics in the 
configuration space on to the stationary points of $V$.
In our glass the particles interact
via a Lennard-Jones potential modified in such a way as to
avoid crystallization \cite{gcr}.
The system is brought in an off-equilibrium initial condition by
a sudden isothermal density jump ({\it crunch}) 
across the liquid-glass transition line. 
%The subsequent evolution of the system is generated by means of
%standard isothermal Molecular Dynamics (MD) technique. We mainly
%focused on the topological description of the energy landscape 
%regions explored by the representative point in configuration space 
%during the off-equilibrium dynamics.
%To this end we partitioned the configuration space in basins of attraction
%of both minima and general stationary points, associating
%to each MD configuration a local minimum IS
%and an {\it Inherent Structure Saddle} (IS-S).     
%The definition of the IS-S may be dependent on the algorithm 
%used to find it, however we expect that the results should be independent
%from the algorithm.
We find a rather sharp cross-over between two
dynamical regimes: 
{\it i}) a short times regime where the system point
travels between basins of saddles of decreasing energy and order and
{\it ii}) a long time regime where the system point moves
between basins of saddles of zero order (minima) and decreasing
energy, crossing basins of saddles of low order.
The existence of two well separated
dynamical regimes shows up also in the Mean Square Displacement (MSD).
Here a logarithmic behavior is found at long times
suggesting the picture of a system point performing a random
walk between basins of minima with a jump probability per unit time 
inversely proportional to time elapsed since crunch.
This picture is quantitatively supported by the very good 
agreement found between the MSD measured along the MD trajectories
and the MSD reconstructed {\it via} a random walk model.

The system under study is a monatomic Modified-Lennard-Jones (MLJ)\cite{gcr}
that is the usual 6-12 Lennard-Jones pair interaction
(truncated and shifted at $R_c=2.6$, standard Lennard-Jones units 
are used hereafter) plus a small
many body term which inhibits crystallization with negligible
corrections to the equation of state of a monatomic
Lennard-Jones system (see \cite{dileo,gcr} for details) . 
The system is composed of $N=256$ particles enclosed in a cubic 
box with periodic boundary conditions.  
The initial liquid equilibrium configurations  
($\rho_0=0.95$, $T_0=0.5$)
are prepared by standard isothermal MD.
At time $t=0$ the density is suddenly increased to the value
$\rho_1=1.24$,
where the glass transition temperature ($T_g(\rho_1)\sim1.4$ \cite{dileo}) 
is well above the simulation temperature $T_0$.
The subsequent evolution of the system 
is generated up to time $t=3\cdot 10^4$ by means of standard
microcanonical MD with time step $\tau=0.01$.
For each MD trajectory we quenched a set of logarithmically equally
spaced configurations to the corresponding IS and saddle
following the steepest descent path respectively on the potential
energy surface $V$ in the former case 
and on the auxiliary potential $W=\frac{1}{2}|\vec{\nabla}V|^2$
\cite{cava0,ange,cava} in the latter.
 
%\vspace{.5cm}
\begin{figure}[hbt]
%\centering
\includegraphics[width=.29\textwidth,angle=-90]{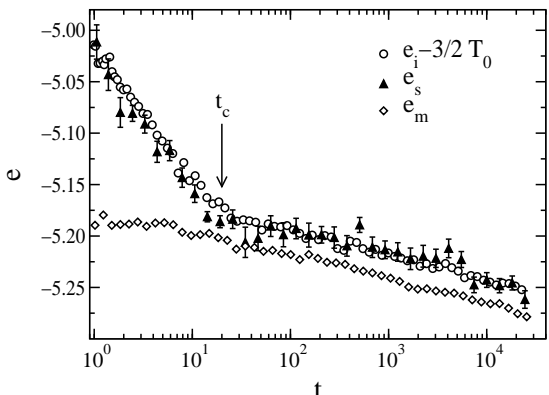}
\caption{
Energies as a function of time during off equilibrium dynamics.
Instantaneous configuration potential energy per particle $e_i$
subtracted by the constant term $\frac{3}{2}T_0$ (open circles),
inherent structure saddle energy $e_s$ (full triangle) and the 
inherent structure minimum  energy $e_m$ (open diamonds).
Error bars are shown only when larger than symbol size.
The arrow indicates time $t_c$ where the crossover from
dynamics between basins of saddles to dynamics between basins of
minima takes place.}
\label{fig1}
\end{figure}

In Fig.\ref{fig1} we report in a logarithmic scale the time evolution of: 
a) the instantaneous configuration potential energy per particle $e_i$ 
subtracted of the constant term $\frac{3}{2}T_0$ (open circles),
b) the saddle's energy $e_s$ (full triangles), 
c) the IS energy $e_m$ (open diamonds).
The remarkable coincidence between $e_s$ and $e_i-\frac{3}{2}T_0$
suggests to think of the instantaneous energy $e_i$ as the sum of two
contributions: the saddle's energy (slowly approaching equilibrium) and
the energy of nearly $N$ fast harmonic equilibrated degrees 
of freedom.  
It is evident the presence of a rather sharp transition in dynamical 
behavior of $e_i$ and $e_s$ at time $t_c\sim20$.
In particular the time evolution of $e_i$ for $t>t_c$ seems to be driven
by the inherent dynamics between minima  as suggested by the 
underlying time evolution of $e_m$.

\begin{figure}[t]
%\centering
\hspace{-.3cm}
\includegraphics[width=.29\textwidth,angle=-90]{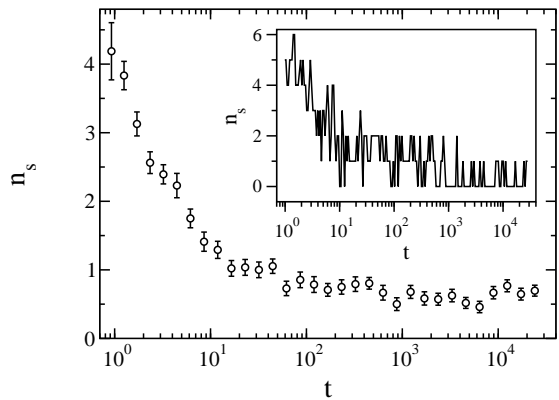}
\vspace{.5cm}
\caption{Mean inherent saddle order $n_s$ (number of negative eigenvalues
of the Hessian matrix of potential energy evaluated at the inherent structure
saddle) as a function of time t after crunch. The inset shows $n_s$ computed 
along a single trajectory.}
\label{fig2}
\end{figure}

In Fig.\ref{fig2} the mean saddle's order $n_s$ 
(number of negative eigenvalues 
of the potential energy Hessian computed on the saddle configurations)
is reported as a function of time. 
The dynamic transition observed at $t_c$ in the energy time evolution 
in Fig.\ref{fig1} corresponds here in the mean order $n_s$ falling below
the value $n_s=1$ or, in other words, for times longer than $t_c$
the system point starts to explore extensively the basins  of saddles
of order zero (minima). The inset in Fig.\ref{fig2} shows the order
$n_s$ computed along a single MD trajectory. One can see that for times 
$t>t_c$ the systems visits basins of minima passing through basins
of low order saddles.
 %\vspace{.52cm}
 
In a recent study of the MLJ landscape probed in equilibrium
\cite{ange} some of us found that the mean elevation energy of a saddle
of order $n_s$ from the underlying minimum is proportional to
$n_s$ (this is at variance with the result reported in \cite{cava},
where a linear relation is observed between $n_s$ and the {\it absolute} 
energy of the saddles). 
This is clearly a property of the energy landscape itself and
one expects that the same properties are also valid in the off-equilibrium dynamics.

\begin{figure}[t]
%\centering
\includegraphics[width=.29\textwidth,angle=-90]{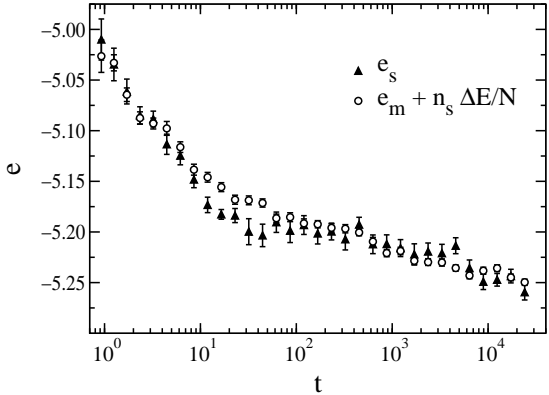}
\vspace{.5cm}
\caption{Saddle's energy $e_s$ (solid symbols) evaluated during off equilibrium dynamics
compared with the energy of the minima $e_m$ increased by the quantity 
$n_s \Delta E/N$ (open symbols) where $n_s$ is the time dependent saddle's order and 
$\Delta E = 10$ is a fitting constant  representing the mean energy separation
between a saddle of order $n_s+1$ and a saddle of order $n_s$.}
\label{fig3}
\end{figure}

In Fig.\ref{fig3} where we report
$e_s$ together with $e_m+n_s \Delta E/N$ with $\Delta E$ a fit parameter.
This result confirms the picture of an energy landscape where the 
saddles of order $n_s+1$ lay above the saddles of order 
$n_s$ of a constant quantity $\Delta E$.
From the data we obtain the value $\Delta E = 10$.
The value of $\Delta E$ could also be obtained from an equilibrium 
analysis, as in \cite{ange} where a value of $\Delta E = 3.6$
is found for the case $\rho=1$. 
We have not carried out such an equilibrium analysis
for the density $\rho=1.24$ used here.  

\begin{figure}[t]
%\centering
%\vspace{.44cm}
\hspace{-.5cm}
\includegraphics[width=.29\textwidth,angle=-90]{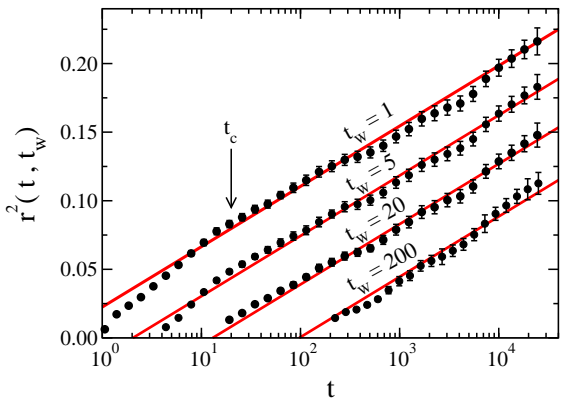}
\vspace{.5cm}
\caption{Mean square displacement $r^2(t,t_w)$ as a function of time $t$ 
for four selected $t_w$ values:
$1, 5, 20$ and $200$ (solid symbols). The straight lines are the prediction of
a random walk model between minima with jump probability per unit time decreasing 
as the inverse of time, $p(t)=c/t$.
According to the model the slope of these lines are $d_0^2 c=0.019$ 
where $d_0^2$ is the mean square distance between
adjacent minima (see text below for details).
The arrow indicates time $t_c$ where the crossover from
dynamics between basins of saddles to dynamics between basins of
minima takes place.
The system size is $N=256$.}
\label{fig4}
\end{figure}
Another interesting quantity in describing the dynamical processes
is the mean square displacement of the configuration at time $t$ from 
that at time $t_w$ after crunch:
\begin{equation}
r^2(t,t_w)=\frac{1}{N}\sum_i\langle |{\mathbf r}_i(t)-{\mathbf r}_i(t_w)|^2\rangle \ 
,
\end{equation}
where $\langle...\rangle$ 
is an average over initial configurations.
In Fig.\ref{fig4} we report $r^2(t,t_w)$ as a function of time $t$ for four
selected $t_w$ values: $t_w=1,5,20,200$. Also in this case $t_c\sim 20$ marks a 
transition in the dynamical behavior. 
In particular the long time regime shows a well defined logarithmic 
dependence on time. 
This suggests the picture of the diffusion process proceeding through elementary 
uncorrelated events occurring with a probability per unit time 
$p(t)\propto 1/t$ \cite{par1}. 
As previously observed, in the long time regime the system point travels
between basins of minima. It is therefore natural to identify the 
elementary diffusion events as jumps between adjacent basins of minima.
In order to give further support to this conjecture we calculated the
cumulative number of jumps $N_{jumps}(t)$ defined as the number of jump
events in which the constantly monitored energy of the IS makes
a sharp transition.

\begin{figure}[tbh]
\centering
\vspace{.5cm}
\hspace{-.5cm}
\includegraphics[width=.35\textwidth]{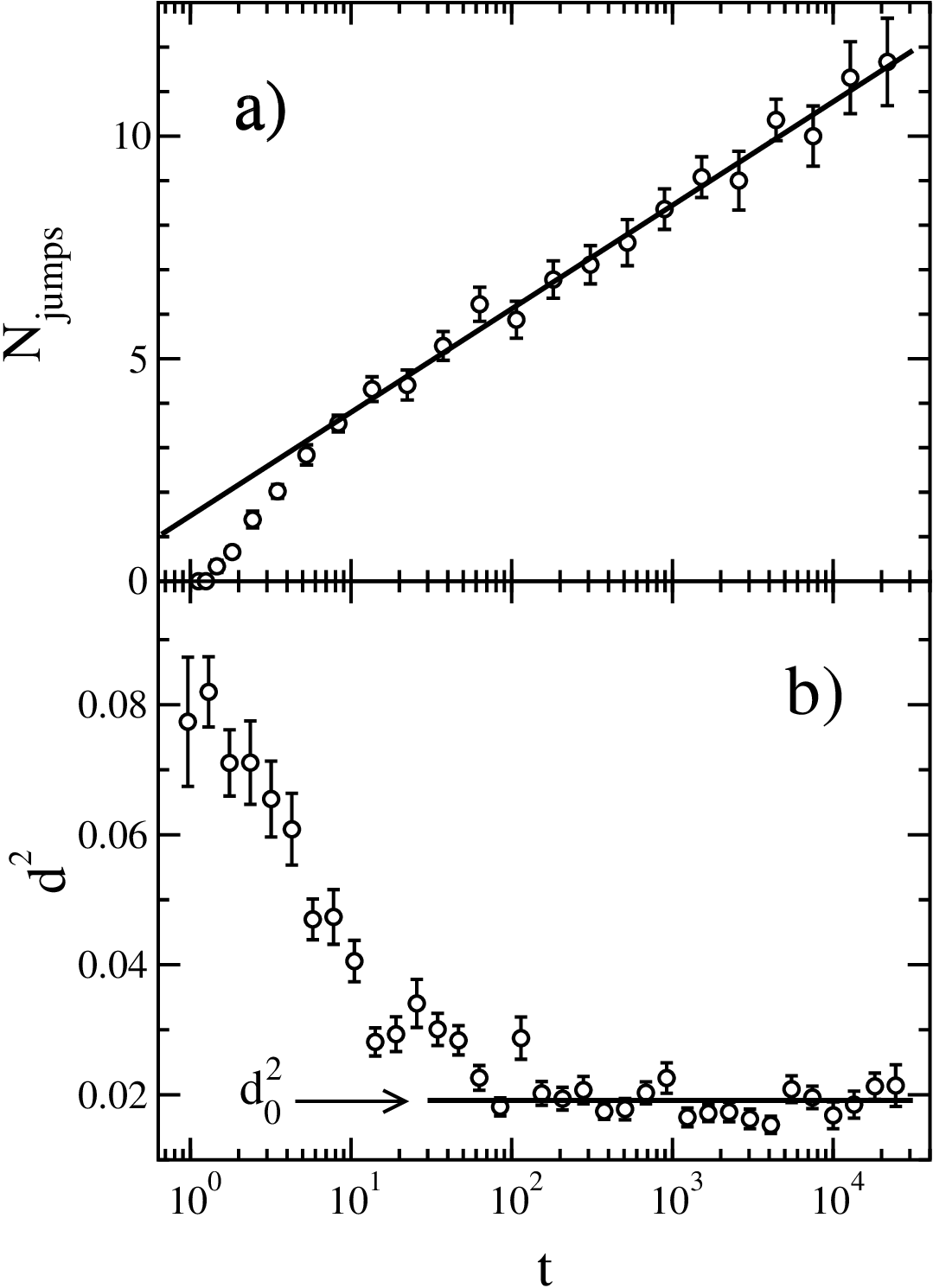}
\vspace{.5cm}
\caption{a) Cumulative number of jumps between adjacent minima visited
by the system during the off-equilibrium dynamics; the solid line is
the best fit for $t>20$ with slope $c=1.0$. 
b)  Mean square distance between adjacent minima $d^2$; the solid
line is the constant value $d_0^2\sim 0.019$ reached for times
greater than $t_c\sim 20$.
The system size is $N=256$. }
\label{fig5}
\end{figure}

The quantity $N_{jumps}$ is plotted in Fig.\ref{fig5}(a).
For times $t>t_c$ $N_{jumps}$ displays a nice logarithmic behavior
that is what one would expect if the probability per unit time of
a jump event is $p(t)=c/t$. In other words the time needed to escape
from an IS reached by the system after a time $t_w$ from {\it crunch} is
proportional to time $t_w$ itself (a similar behavior is found in the
off-equilibrium dynamics of a ferromagnetic Ising system where the persistence
time of a bubble in which spins have a similar value is proportional to
time $t_w$ \cite{par2,bubble}).
In Fig.\ref{fig5}(a) the solid line
represents the fitting function $c\log(t)$ with $c=1.0$. 
%%%%%%%%%%%%%%%%%%%%%%%%%%%%%%%%%%%%%%%%%%%%%%%%%%%%
In Fig.\ref{fig5}(b) we report the mean square distance $d^2$ between 
adjacent minima probed by the system during the aging dynamics:
\begin{equation}
d^2 (t) =\frac{1}{N}\sum_i\langle 
|{\mathbf r}_i^{IS}(t^+)-{\mathbf r}_i^{IS}(t^-)|^2\rangle \ ,
\end{equation} 
where $t$ is the time of the jump between two different basins of minima
and ${\mathbf r}_i^{IS}(t^-)$, ${\mathbf r}_i^{IS}(t^+)$ are
the coordinates of IS  respectively before and after time $t$. 
The quantity reported in figure approaches the long time constant 
value $d_0^2 \sim 0.019$.
%%%%%%%%%%%%%%%%%%%%%%%%%%%%%%%%%%%%%%%%%%%%%%%%%%%%
%On the other hand the mean square distance between adjacent minima $d^2$,
%reported in Fig.\ref{fig5}(b), approaches the constant value $\sim 0.019$.
For uncorrelated jumps, the mean square displacement
would be simply given by 
\begin{equation}
r^2(t,t_w)=d_0^2 c \log(t/t_w) \ .
\end{equation}
This is in a remarkably good agreement with the MD data as shown
in Fig.\ref{fig4}, where  we draw straight lines with slope $d_0^2 c=0.019$ 
(as obtained from the values of $d_0^2$ and $c$ derived from Fig.\ref{fig5})
and intercept chosen as to best fit the data for $t>20$.

In conclusion we have numerically studied the off-equilibrium 
dynamics of a monatomic Modified-Lennard-Jones system by sudden 
isothermal density jumps from the liquid to the glassy phase.
Analyzing the time evolution of the energy and the mean square 
displacement we found two well defined 
dynamical regimes, which can be related to the different ways 
the system explores the potential energy landscape:
{\it i}) at short times the system point moves downward passing
through basins of saddles with decreasing energy and order, and 
{\it ii}) at longer time the relevant dynamical processes are 
jumps between basins of minima with slower decreasing energy.
In the whole time regime the instantaneous potential energy $e_i$ 
can be thought as due to a sum of two contributions: 
the energy of the underlying saddle's configurations $e_s$ plus a 
nearly harmonic term of the fast degrees of freedom
$3/2 T$. 
Moreover the linear energy gap between saddles and underlying minima
$e_s-e_m$ as a function of saddles order $n_s$ supports 
the picture of a simple energy landscape with only a single energy barrier
parameter $\Delta E$ which represents the energy gap between a saddle 
of order $n_s$ and a saddle of order $n_s+1$.
The relevant processes in the long time aging regime are well represented
by a simple random walk model with elementary events  
corresponding to jumps between minima with a probability per unit time 
proportional to the  inverse of time elapsed since the system has been
brought out of equilibrium.

%%%%%%%%%%%%%%%%%%%%%%%%%%%%%%%%%%%%%%%%%%%%%%%%%%%%%%%%%%%%%%%%%%%%%%%%%%%
%                             REFERENCES
%%%%%%%%%%%%%%%%%%%%%%%%%%%%%%%%%%%%%%%%%%%%%%%%%%%%%%%%%%%%%%%%%%%%%%%%%%%

\end{document}